# Magnetic correlations in spin ice $Ho_{2-x}Y_xTi_2O_7$ as revealed by neutron polarization analysis

L. J. Chang[1,2*], Y. Su[3†], Y. -J. Kao[4‡], Y. Z. Chou[4], R. Mittal[3,5], H. Schneider[3], Th. Brückel[3,6], G. Balakrishnan[7], and M. R. Lees[7]

[1] *Nuclear Science and Technology Development Center, National Tsing Hua University, Hsinchu 30013, Taiwan*

[2] *Quantum Beam Science Directorate, Japan Atomic Energy Agency (JAEA), Tokai, Ibaraki 319-1195, Japan*

[3] *Jülich Centre for Neutron Science (JCNS), IFF, Forschungszentrum Jülich, Outstation at FRM-II, Lichtenbergstrasse 1, D-85747 Garching, Germany*

[4] *Department of Physics and Center of Quantum Science and Engineering, National Taiwan University, Taipei 10607, Taiwan*

[5] *Solid State Physics Division, Bhabha Atomic Research Center, Trombay, Mumbai 400 085, India*

[6] *Institut für Festkörperforschung, Forschungszentrum Jülich, 52425 Jülich, Germany*

[7] *Department of Physics, University of Warwick, Coventry CV4 7AL, United Kingdom*

**\*Corresponding author:** liehjeng.chang@jaea.go.jp

[†] E-mail address: y.su@fz-juelich.de

[‡] E-mail address: yjkao@phys.ntu.edu.tw

We present single-crystal magnetic diffuse neutron-scattering measurements with polarization analysis on the spin ice $Ho_{2-x}Y_xTi_2O_7$ ($x$ = 0, 0.3, and 1). At 2 K, the spin flip scattering patterns are typical of a nearest-neighbor spin ice and the effects of Y doping are seen in the widths of the pinch points. At low temperature, the spin flip patterns are characteristic of a dipolar spin ice and are unaffected by Y doping. In the non-spin flip channel, we observe the signature of strong antiferromagnetic correlations at the same temperature as the dipolar spin ice correlations appear in the spin flip channel. These experimental observations have been well reproduced by our Monte Carlo simulations.

Pyrochlore oxides, with chemical formula $A_2B_2O_7$, where A is a rare-earth ion and B is a transition-metal ion, have attracted great attention both experimentally and theoretically.[1] Pyrochlore oxides have a face-centred-cubic structure with space group $Fd\bar{3}m$ and eight molecules of its general formula in a unit cell.[1,2] Their unusual structure of corner-sharing tetrahedra leads to geometrical frustration and exotic low-temperature magnetic properties[2] such as a collective paramagnet state,[3] spin glass behavior,[4] and the most extensively studied dipolar spin ice.[5-9] The spin ice compounds $Ho_2Ti_2O_7$, $Ho_2Sn_2O_7$, $Dy_2Ti_2O_7$, and the latest member of the family, $Dy_2Sn_2O_7$,[10] in which only the rare earth elements possess magnetic moments, exhibit a local <1,1,1> Ising-like anisotropy and an effective ferromagnetic interaction between the magnetic moments.[8-14] The organizing principles of the magnetic ground-state, or the "ice rules", require that two spins should point in, and two out of each elementary tetrahedron in the pyrochlore lattice. The minimal model for the system is the dipolar spin ice model (DSM) with the nearest-neighbor exchange and long-range magnetic dipole interactions, given by the Hamiltonian

$$H = -J\sum_{\langle ij \rangle} \mathbf{S}_i \cdot \mathbf{S}_j + Da^3 \sum_{i>j} \left[ \frac{\mathbf{S}_i \cdot \mathbf{S}_j}{r_{ij}^3} - \frac{3(\mathbf{S}_i \cdot \mathbf{r}_{ij})(\mathbf{S}_j \cdot \mathbf{r}_{ij})}{r_{ij}^5} \right], \quad (1)$$

where $a$ is the nearest neighbor distance, $\mathbf{S}_i = \sigma_i \hat{z}_i$ is a spin of unit length along the <1,1,1> Ising orientation $\hat{z}_i$.[11,15-17] At high temperatures, the low energy physics can be adequately described by the nearest-neighbor spin ice model (NNSM), which is the DSM Hamiltonian truncated at the nearest-neighbor interactions, as a result of the self-screening of the dipolar interactions on the pyrochlore lattice.[18] The magnetic correlations have the spatial dependence of a dipolar form, resulting in characteristic pinch-point features in the neutron scattering data.[19] The dipolar spin ice ground-state manifold is, however, quasi-degenerate, and the residual energy scale leads to an ordering instability at low-$T$ away from $\mathbf{q} = \mathbf{0}$,[15-16] though this long-range ordered state has not yet been observed experimentally. Excitations above the ground state manifold, which locally violate the ice rules, can be viewed as magnetic monopoles of opposite "charge" connected by Dirac strings.[20] Evidence of magnetic monopoles in spin ice has been observed in several recent experiments.[21-25] In particular, Fennell *et al.* recently studied the magnetic correlations in $Ho_2Ti_2O_7$ in the spin flip (SF) and the non-spin flip (NSF) channels via polarized neutron scattering. While the scattering pattern of a nearest-neighbor spin ice model was clearly observed in the SF channel at 1.7 K, a faint indication of the presence of antiferromagnetic (AFM) correlations was found in the NSF channel.[21]

One might expect that impurity doping would relax the local constraints imposed by the ice rules, and that significant changes would be observed in the doped spin ices. However, studies of doped spin ice by replacing Ho or Dy with Y, and stuffed spin ice materials, suggest that the spin ice state is present even when the amount of doping is as high as 40%.[14] Relaxation measurements by neutron spin-echo and ac-susceptibility showed that the relaxation mechanism in spin-ice compounds is also less sensitive than expected to the concentration of the dopants.[26,27] Despite much work in this area, several outstanding questions remain. For instance, what is the effect of doping on the magnetic correlations in the dipolar spin ice, given that other characteristics of the spin ice persist even at large chemical doping?[14,26,27] Is there any experimental signature of long-range order in these materials? What the temperature regime where the NNSM fails to describe the magnetic correlations in the system?

We present a comprehensive investigation of the magnetic diffuse scattering in the (h, h, l) reciprocal plane for single crystals of $Ho_{2-x}Y_xTi_2O_7$ ($x$ = 0, 0.3, and 1) via neutron polarization analysis to study the short-range magnetic correlations in these materials. In the neutron SF scattering, we observe a crossover at $T \sim 2$ K in the diffuse scattering from patterns that are typical of a NNSM at high-temperature to those representative of a DSM at low temperature.[28] In the NSF channel, diffuse scattering corresponding to apparent short-range AFM correlations is observed at the zone boundaries below 2 K. We argue that the AFM correlations in the NSF channel are the precursor to the illusive long-range ordered ground state in the dipolar spin ice predicted in Refs. 15, and 16. Surprisingly, at low-$T$ no significant changes in the magnetic scattering patterns can be found as the system is doped with Y. The observed magnetic diffuse scattering patterns can be well reproduced by our Monte Carlo simulations of the system.

$Ho_{2-x}Y_xTi_2O_7$ ($x$ = 0, 0.3, and 1) single crystals were grown in an infrared double mirror image furnace.[29] Neutron scattering experiments were performed at the high-flux polarized diffuse neutron scattering spectrometer DNS, FRM-II (Garching, Germany). A $^3$He/$^4$He dilution refrigerator insert with an Oxford Instruments cryostat was used for temperatures below 2 K, and a closed cycle cryostat was used above 2 K. A neutron wavelength of 4.74 Å was chosen for all the experiments. The [1, -1, 0] direction of the crystals was aligned perpendicular to the horizontal scattering plane so that the (h, h, l) reciprocal plane can be mapped out by rotating the samples. The neutron polarization at the sample position was aligned along the [1, -1, 0] direction of the sample, i.e. the $z$-direction of the chosen experimental coordinate system. Within this setup, the SF and NSF scattering cross sections are $\left(\frac{d\sigma}{d\Omega}\right)^{SF}_z = M^*_{\perp y}M_{\perp y} + \frac{2}{3}I_{SI}$ and

$$\left(\frac{d\sigma}{d\Omega}\right)_z^{NSF} = M_{\perp z}^* M_{\perp z} + N^* N + \frac{1}{3} I_{SI},$$ where $M_{\perp y}^* M_{\perp y}$ and $M_{\perp z}^* M_{\perp z}$ are the components of the magnetic scattering cross section in and out of the (h, h, l) scattering plane respectively. $I_{SI}$ is the total nuclear spin incoherent scattering cross section, and $N^* N$ is the nuclear coherent scattering cross section.[30]

We have modeled the observed magnetic diffuse scattering patterns using Monte Carlo simulations of the dipolar spin ice model. In these simulations, the nearest-neighbor exchange, $J$, for $Ho_2Ti_2O_7$ is $J$ = -1.56 K, and the dipolar coupling, $D$, is $D = \frac{\mu_0}{4\pi}\frac{\mu^2}{a^3}$ = 1.41 K in the DSM Hamiltonian Eq. (1). This corresponds to an effective nearest-neighbor exchange $J_{nn}$ = J/3 + 5D/3 = 1.83 K. We simulate the Hamiltonian using a Metropolis Monte Carlo algorithm combining single spin and loop dynamics.[17] The long-range dipolar interactions are handled using Ewald summation techniques.[31] The simulations were performed on three-dimensional lattices of $L \times L \times L$ cubic cells of the pyrochlore lattice under periodic boundary conditions, with $L$ = 4, 6, 8, and the total number of spins in the simulation cell was $N = 16 \times L \times L \times L$. The simulations contain 10000 non-local loop updates and single spin flip per spin for both equilibrium and sampling processes. Each loop is generated by flipping connected spins until a closed path is found, and the MC dynamics is reminiscent of the creation and annihilation of a pair of magnetic monopoles with opposite charges.[17,20] We collected data from 50-500 disorder realizations for each of the doped cases. For each disorder realization, we calculate the Fourier transformation of the real space correlations, $I(\mathbf{q})$, by averaging over 256 different spin configurations. $I(\mathbf{q}) \propto \frac{|f(\mathbf{q})|^2}{N} \sum_{ij} \langle s_i s_j \rangle (\hat{z}_i^\perp \cdot \hat{z}_j^\perp) e^{i\mathbf{q}\cdot\mathbf{r}_{ij}}$ denotes thermally averaged correlations between the Ising spins at sites $i, j$; $\hat{z}_i^\perp$ is the component of the quantization direction at site i perpendicular to the scattering vector $\mathbf{q}$, $N$ is the number of spins and $f(\mathbf{q})$ is the magnetic form factor of $Ho^{3+}$.[32]

Figure 1 shows the experimental SF scattering patterns for the $Ho_{2-x}Y_xTi_2O_7$ with (a) $x$ = 0.3 at 400 mK, (b) $x$ = 1.0 at 30 mK, and (c) $x$ = 0.3 at 2 K. The residual intensities at the nuclear peak positions in the SF scattering are due to the finite neutron polarization rate of the incident neutron beams (~95% throughout the experiments). The low-T results ((a) and (b)) show clearly a distinct DSM scattering pattern with four intense regions around (0, 0, 0), and the spread of the broad features connecting (0, 0, 0) to regions of intensity around (0, 0, 3) and (3/2, 3/2, 3/2).[10,28] The pinch points, which are associated with dipolar correlations in real space, are clearly observed at the (0, 0, 2) position in Fig. 1 (a) and (b) as bow-tie shapes.

One might expect that the ice rules are relaxed due to the vacancies introduced by Y in each tetrahedron, thus weakening the spin-ice correlations. The experimental results, however, show that the Y dilution does not alter the scattering patterns significantly even with $x = 1.0$, i.e., half of the spins removed. This result agrees with previous observations that the spin ice state is robust against dilution. One possible explanation for this peculiar behavior is that the low energy excitations in the spin ice, for either the clean or the doped systems, involve the collective flipping of clusters of spins. Although the local ice rules might be violated due to impurities, the low-energy spectrum is not changed much by doping, thus no major difference in the scattering is observed between the doped and the undoped systems. On the other hand, if the doping level is increased such that these collective excitations are no longer available, one would expect the system to be described by simple single spin physics.

We also observe a crossover from a magnetic correlation pattern that is typical of the DSM at 400 mK to a pattern that is chatacteristic of the NNSM at 2 K and which persists up to 10 K.[21,28] MC simulations of the DSM at 500 mK (Fig. 1 (e)) and 2 K (Fig. 1 (f)) confirm these results. So, at high-$T$, the energy scale between states in the quasi-degenerate manifold is irresolvable due to the thermal fluctuations, and the scattering can well be described by the NNSM. At low-$T$, the residual energy scales become relevant and it is necessary to include the full dipolar interaction to account for the details of the scattering patterns.[28]

To further investigate the effects of dilution, Fig. 2 shows cuts of the SF scattering crossing the pinch point at (0, 0, 2) along (h, h, 0) at low-$T$ ($x = 0, 1$ at 30 mK, and $x = 0.3$ at 400 mK) and at 2 K. For comparison, we performed simulations of the DSM at 500 mK and the NNSM at $T = 1.3$ K at various doping levels. At low-$T$, the system is described by the dipolar spin-ice model, and the widths of the pinch point are not affected by dilution (Fig. 2 (a) and (c)). At 2 K, the effects of Y dilution are clearly observed in Fig. 2(b). The sensitivity of the correlations to dilution can be also seen in the simulations of the NNSM (Fig. 2(d)). These results show that doping reveals the temperature regimes where the DSM and the NNSM provide a valid description of the system.[18] At 2 K, although the microscopic Hamiltonian is given by Eq. (1), the low-energy physics is described by the NNSM,[21] and doping with Y can effectively destroy the singular behavior at the pinch point. At lower temperature, one needs to take into account the full long-range dipolar interactions as the simpler NNSM description is no longer valid. In this respect, we can understand why the low-$T$ magnetic correlations are insensitive to doping, since the correlations at all distances contribute to the scattering amplitude. One might expect that further dilution may also destroy these correlations, since in the very dilute limit, single spin behavior should be observed. Further studies on higher Y doped compounds are necessary to confirm this assumption.

Figure 3 (a) and (b) show the NSF scattering for $x = 0.3$ at 400 mK, and a MC simulation using the DSM at 500 mK respectively. The pattern in Fig. 3(a) corresponds to the component of the magnetic correlations perpendicular to the (h, h, l) scattering plane, and is only observed when the dipolar spin ice pattern appears in the SF channel. Signatures of strong AFM correlations along (0, 0, l) and (h, h, 0) are observed for all the compounds. One of the possible antiferromagnetic configurations is depicted in the inset of Fig. 3(a).[16] It is noteworthy that the magnetic scattering cross sections in and out of the scattering plane in the experiments are not identical.[21] The observed anisotropic magnetic correlations are likely to result from the Ising anisotropy of the spin ice. When the temperature is increased to ~2 K, where in the SF channel the magnetic correlations cross over to the NNSM-type, only very weak correlations are observed in the NSF channel (Fig. 3(c)). The cuts along (h, h, 2) centered at (1, 1, 2) at low-$T$ are shown in Fig. 3(d), and show very slight doping dependence. Our experimental results provide strong evidence that the AFM correlations in the NSF channel and the quasi-degenerate manifold in the DSM are closely related.[15,16] The emergence of the strong AFM correlations, related to the possible long-range order, suggests that impurity doping may locally relieve the frustration and the spins can relax more easily into the ordered state. For water ice, a first-order transition is induced by a small amount of KOH.[33] In contrast, Y doping in spin ice seems to have little effect on the ordering mechanism, as observed from the peak widths in Fig. 3(d). On the other hand, preliminary calorimetric results indicate impurity doping removes part of the residual entropy of the spin ice state,[34] and possibly induces a phase transition at high doping concentration. Higher Y doped systems may clarify the possible ordering mechanism in these systems.

In conclusion, we have studied the complex magnetic correlations in Y doped single crystals of the spin ice $Ho_2Ti_2O_7$ using diffuse neutron-scattering with polarization analysis. For all the specimens studied, at temperatures between 2 K and 10 K the magnetic correlations produce SF neutron scattering patterns that can be described using a nearest-neighbor spin ice model. Below ~2 K the scattering patterns can be described by a dipolar spin ice model. At low-$T$, the widths of the pinch points are not affected by dilution, while at 2 K, the effects of Y dilution become significant. These results show that doping reveals the regime where the DSM and NNSM are valid descriptions of the system. The signature of AFM correlations, which are observed along the zone boundary in the NSF scattering, and at positions (0, 0, 1), (0, 0, 3), (3/2, 3/2, 3/2) in the SF channel, may be the precursor of the predicted long-range order in the spin ice.[16]

We thank W. Schweika, P. Holdsworth, and B. Gaulin, for valuable discussions, and M. Gingras, and Y. Endoh for critical reading of the manuscript. This work was supported in Taiwan by grants NSC 96-2739-M-213-001 and NSC 99-2112-M-007-020 (LJC), and NSC 97-2628-M-002-011-MY3 and NTU 98R0066-65, -68 (YZC and YJK). Artur Glavic is acknowledged for implementing the *dnsplot* software package.

Figure captions

FIG. 1. (Color online) Experimental neutron SF scattering patterns for the $Ho_{2-x}Y_xTi_2O_7$ compounds for (a) $x = 0.3$ at 400 mK, (b) $x = 1.0$ at 30 mK, (c) $x = 0.3$ at 2 K, and (d) $x = 0.3$ at 10 K; MC simulations made using the DSM for (e) $x = 0.3$ at 500 mK, and (f) $x = 0.3$ at 2 K.

FIG. 2. (Color online) Cuts across the pinch point (0, 0, 2) along (h, h, 0) from the experimental neutron SF scattering for different Y doped compounds (a) at low-$T$ ($x = 0, 1$, $T = 30$ mK, and $x = 0.3$, $T = 400$ mK); (b) at $T = 2$ K; (c) MC simulation using the DSM at 500 mK; and (d) MC simulation using the NNSM at $T = 1.3$ K. The intensities are normalized to the maximum intensity in each set of data. The lines are Lorentzian fits.

FIG. 3. (Color online) (a) NSF scattering for $x = 0.3$ at 400 mK. The inset shows one of the possible antiferromagnetic configurations ($x$-$y$ plane projection and in three dimensions, +/- indicates that the moment is along +$z$/-$z$ direction) predicted by Ref. 16; (b) MC simulation results using the DSM for $x = 0.3$ at 500 mK; (c) NSF scattering for $x = 0.3$ at 2 K; the low-temperature (400 mK) pattern is not observed at this temperature; (d) cuts along (h, h, 2) for different doping levels at low temperature (for $x = 0, 1$, $T = 30$ mK, and $x = 0.3$, $T = 400$ mK). The lines are Lorentzian fits.

FIG. 1

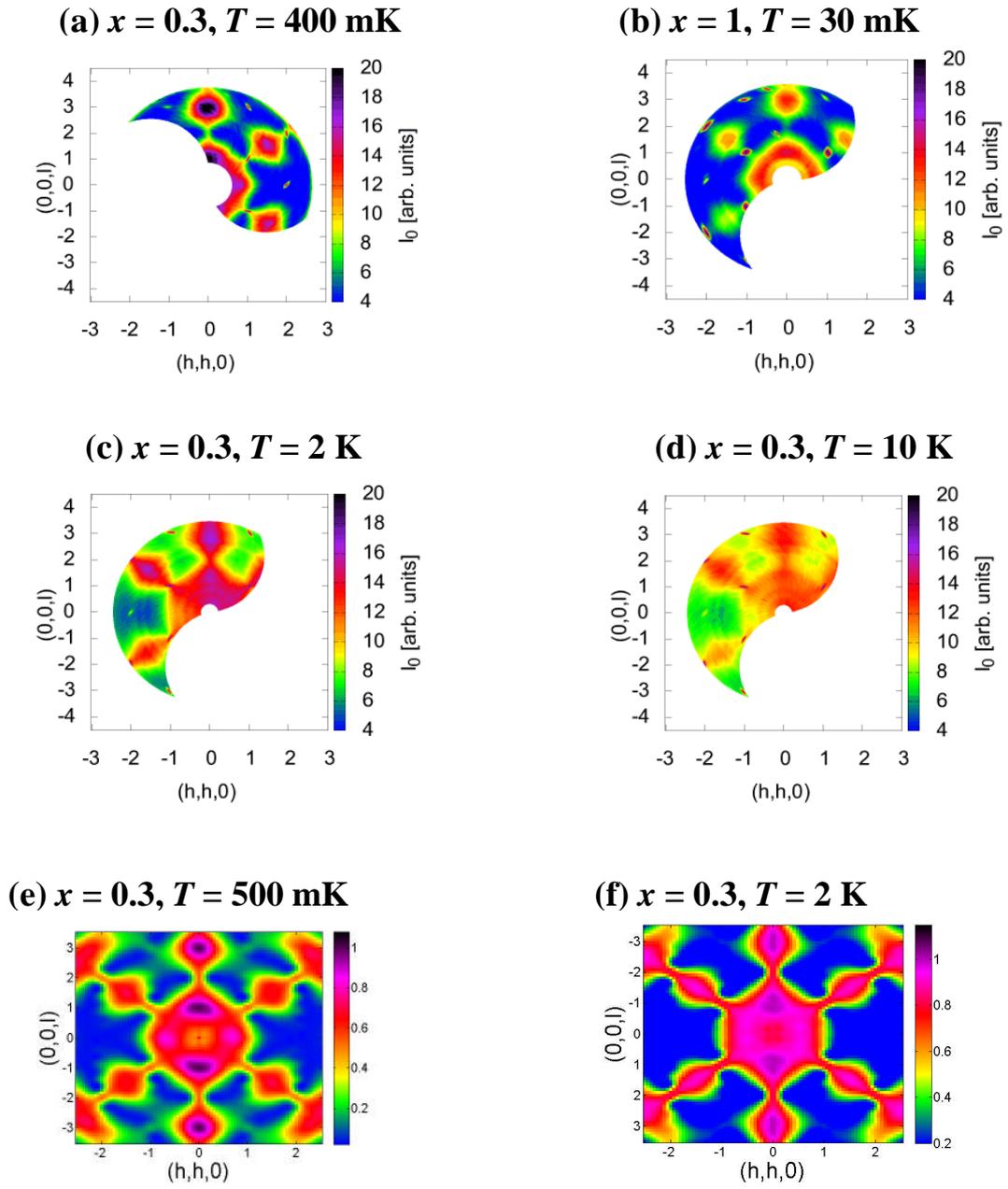

FIG. 2

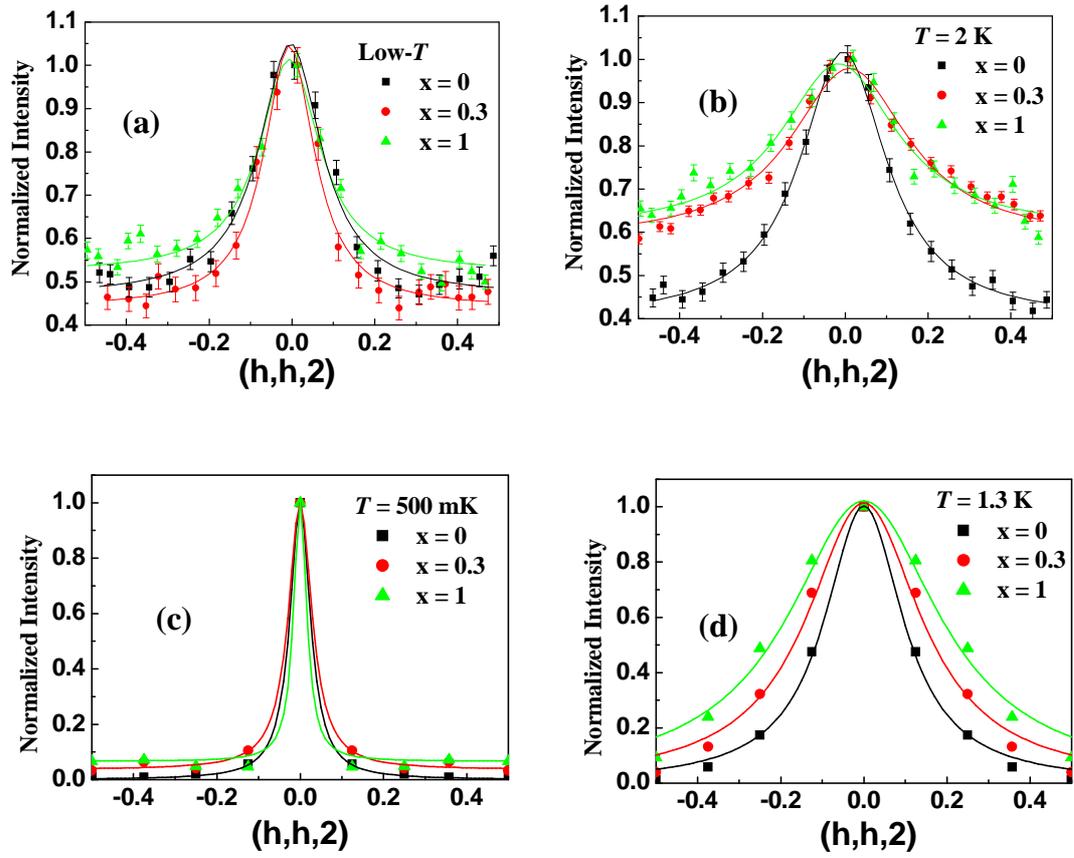

FIG. 3

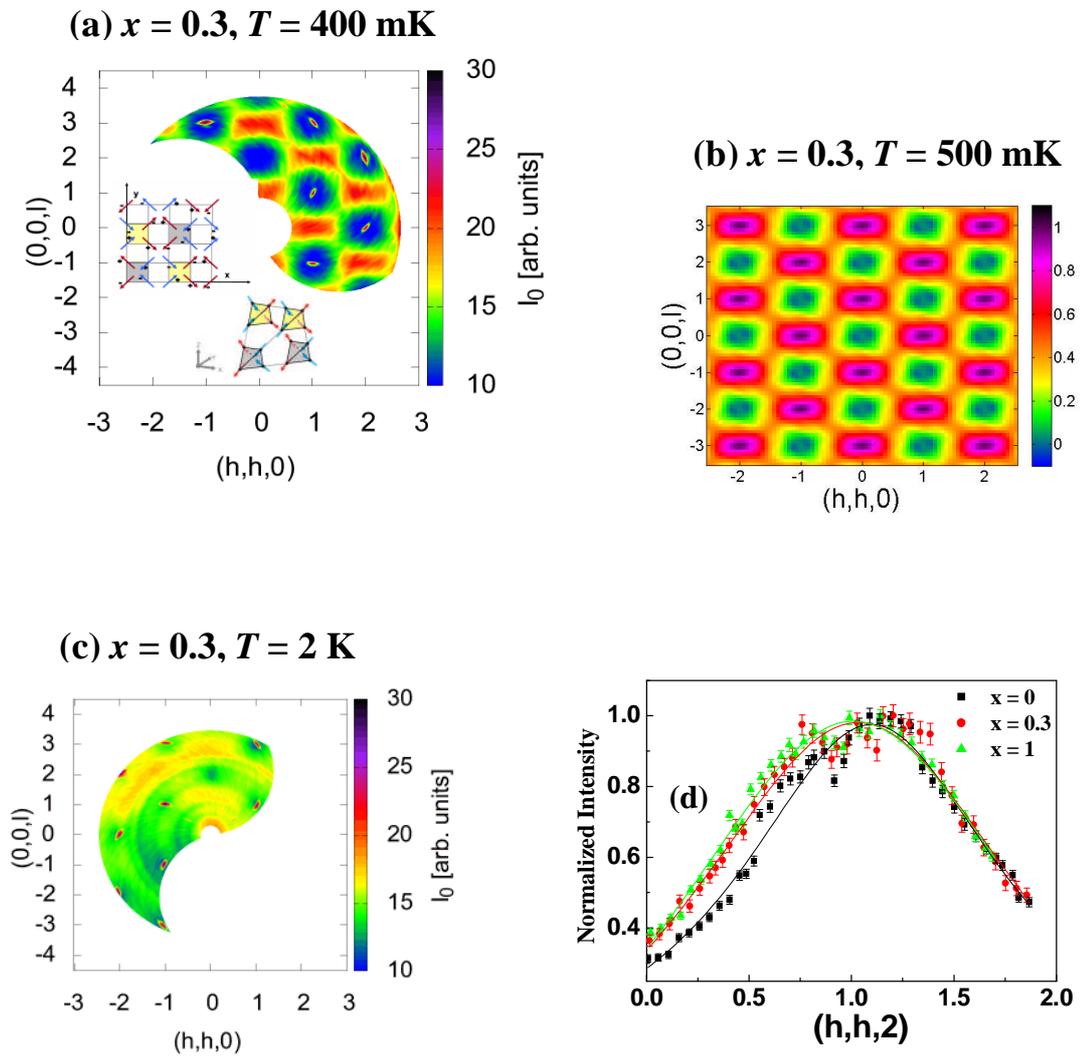